\documentclass{PoS}

\usepackage{BibDef}

\title{Star Cluster Disruption by a Supermassive Black Hole Binary}

\ShortTitle{Star Cluster  Disruption by a SMBH binary}

\author{\speaker{Elisa Bortolas}\\
        INAF, Osservatorio Astronomico di Padova, Vicolo dell'Osservatorio 5, I-35122 Padova, Italy\\
        Dipartimento di Fisica e Astronomia `Galileo Galilei', Universit\`a di Padova, Vicolo dell'Osservatorio 3, I-35122 Padova, Italy\\
        E-mail: \email{elisa.bortolas@phd.unipd.it}}

\author{{Michela Mapelli}\\
        INAF, Osservatorio Astronomico di Padova, vicolo dell'Osservatorio 5, I-35122 Padova, Italy\\
        Institute for Astrophysics and Particle Physics, University of Innsbruck, Technikerstra\ss e 25/8, A-6020 Innsbruck, Austria\\
        INFN, Milano Bicocca, Piazza della Scienza 3, I-20126 Milano, Italy\\
        E-mail: \email{michela.mapelli@oapd.inaf.it}}

\author{{Mario Spera}\\
        INAF, Osservatorio Astronomico di Padova, vicolo dell'Osservatorio 5, I-35122 Padova, Italy\\
        Institute for Astrophysics and Particle Physics, University of Innsbruck, Technikerstra\ss e 25/8, A-6020 Innsbruck, Austria\\
        INFN, Milano Bicocca, Piazza della Scienza 3, I-20126 Milano, Italy\\
        E-mail: \email{mario.spera@uibk.ac.at}}

\abstract{%
Supermassive black hole binaries (BHBs) are expected to be one of the most powerful
sources of low-frequency gravitational waves (GWs) for future space-borne detectors.
Prior to the GW emission stage, BHBs evolving in gas-poor nuclei shrink primarily through the slingshot ejection of stars approaching the BHB from sufficiently close distances. Here we  address the possibility that the BHB shrinking rate is  enhanced through the infall of a star cluster (SC) onto the BHB. 
We present the results of direct summation $N$-body simulations exploring different orbits for the SC infall, and we show that SCs reaching the
BHB on non-zero angular momentum orbits (with eccentricity 0.75) fail to enhance the
BHB hardening, while SCs approaching the BHB on radial orbits reduce the BHB separation by $\sim 10\%$ in less than 10 Myr, effectively shortening the BHB path towards GWs.}

\FullConference{GRAvitational-waves Science\& technology Symposium - GRASS2018\\
		1-2 March 2018\\
		Palazzo Moroni, Padova (Italy)}

\begin{document}

\section{Introduction}
Supermassive black hole binaries (BHBs) are a natural by-product of galaxy mergers; as such, they are expected to form in large numbers along cosmic time \cite{Begelman1980}.
The study of BHBs has received  considerable  attention in the last decades: when the BHB members reach separations of the order of a few mpc, they coalesce into a single supermassive black hole (SMBH) via a burst of low-frequency gravitational waves (GWs) \cite{ThorneBraginskii1976}. Such GW sources  will shine in the band of the future space-borne LISA observatory, making BHB mergers among the main targets for the LISA mission \cite{Amaro-Seoane2017}.

In the early stages of a galaxy merger, dynamical friction drives the SMBHs toward the centre of their
common potential well.  When the BHB separation drops below parsec scales, the BHB shrinking in gas-poor nuclei  is primarily driven by three body scatterings (`slingshot ejections') of stars on sufficiently low angular momentum orbits \cite{Begelman1980}.
At the beginning of the slingshot phase, the BHB
promptly expels from the galaxy almost all stars able to reach its neighbourhood. After that, the BHB
shrinking can slow down considerably and even stop at roughly parsec scale, unless any physical processes can guarantee a
steady repopulation of the binary loss cone (i.e. the region of phase-space containing stars with sufficiently low angular momentum) \cite{Milosav2003}. For this reason, a number of studies in the last decades have investigated whether BHBs can efficiently merge in gas-poor environments  \cite{Begelman1980,Bortolas2016}.  Currently, a general solution to the `final parsec problem' is believed to reside in the non-sphericity of the host galaxy \cite[and references therein]{Khan2016,Gualandris2017}. If the host galaxy is triaxial, gravitational torques ensure a steady scattering of stars into the BHB loss cone \cite{Yu2002}; given that galaxy mergers are expected to naturally induce non-sphericity in the merger relic \cite{Bortolas2018}, most BHBs are believed to find their way to coalescence within a few  Gyr \cite{Khan2016}. 

Here we propose a novel possible way to shorten the BHB path to coalescence, via the infall of a massive  stellar cluster (SC) onto a parsec-scale BHB. In fact, young SCs are common in
galactic nuclei, and may form in a burst of star formation following the galaxy merger \cite{Sanders1988}; then, SCs may sink towards the centre of the system via dynamical friction,  interacting with the BHB, and possibly significantly contributing to its hardening.
The work presented here is a summary of \cite{Bortolas2018a}; a further study  addressing the effect of SC infalls onto BHBs can be found at \cite{Arca-Sedda2017}.

\section{Methods and initial conditions }
We simulate the infall of a SC onto a BHB adopting the highly accurate, direct summation $N$-body code HiGPUs \cite{Spera2013}. 
We place two  $10^6 {\rm M}_\odot$ SMBHs on a circular orbit with semimajor axis $a=1$ pc; the BHB centre of mass coincides with the bottom of the galactic potential well, modelled as a rigid Dehnen profile \cite{Dehnen1993} with total mass $5\times10^{10}{\rm M}_\odot$, scale radius 250 pc, and inner density slope $\gamma=0.5$. 
A $\approx 8\times 10^4 {\rm M}_\odot$ SC is set at 20 pc from the BHB centre of mass; the SC follows  an isotropic King density profile \cite{King1966} with core radius $r_k= 0.4$ pc, and the SC mass spectrum is modelled with a Kroupa mass function \cite{Kroupa2001}. 
We explore three different configurations for the cluster infall: in run 1p (2p), the SC is initially at rest, and its orbital plane is perpendicular (coplanar) with respect to the BHB orbital plane;
in run 3p, the SC is placed at the apoapsis of
an eccentric orbit with eccentricity 0.75, coplanar with the BHB, with counter-aligned angular momentum. Each simulation is integrated for 10 Myr.
%
%Below we describe the interaction between the SC and the SMBHB.

\section{Results}
\begin{figure}[bt]
	\includegraphics[width=\textwidth]{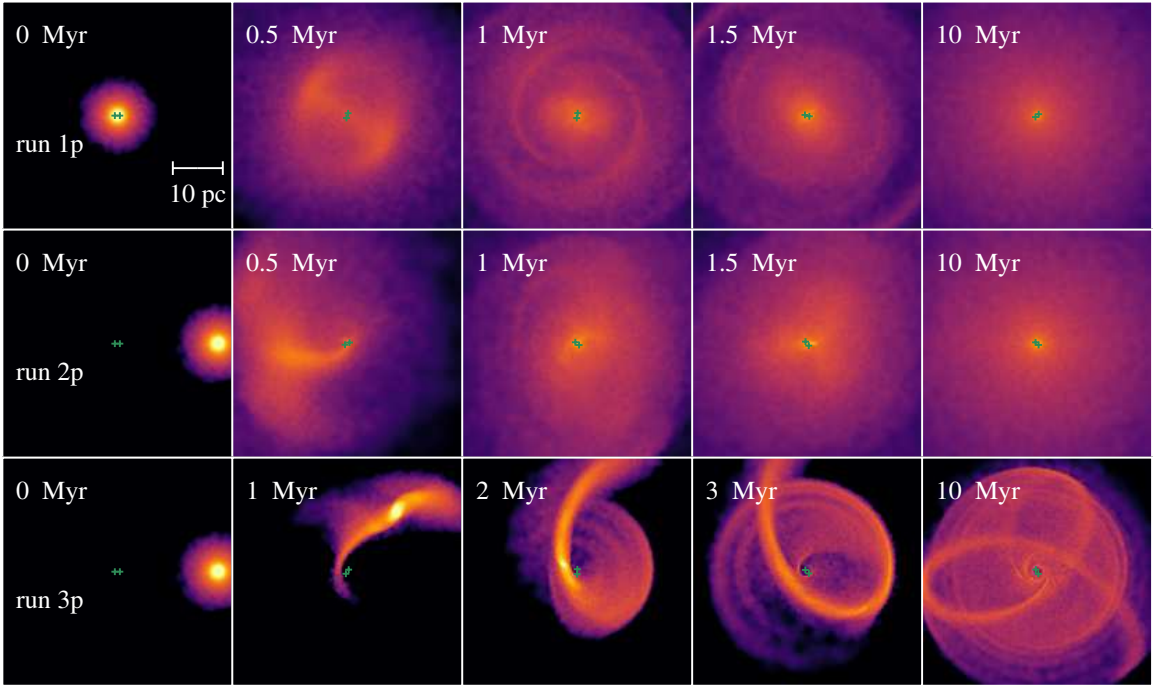} 
		\caption{ Time evolution of the stellar surface density projected on the BHB orbital plane for runs~1p (top row), 2p (central row), 3p (bottom row). The green central crosses mark the position of the two SMBHs. The colour coding is logarithmic, and it  refers to the smoothed projected mass density of stars, ranging from $10^{-1}$ (black) to $10^4$ M$_\odot$ pc$^{-2}$ (white).}
    \label{fig:snap}
\end{figure}

Fig.~\ref{fig:snap} shows different snapshots of the simulations projected on the BHB orbital plane. The BHB response to the SC infall significantly depends on the SC orbital angular momentum.

\subsection{Radially infalling SCs}
\begin{figure}
	\includegraphics[trim={10cm 0cm 2cm 16cm},angle=270,width=0.5\textwidth]{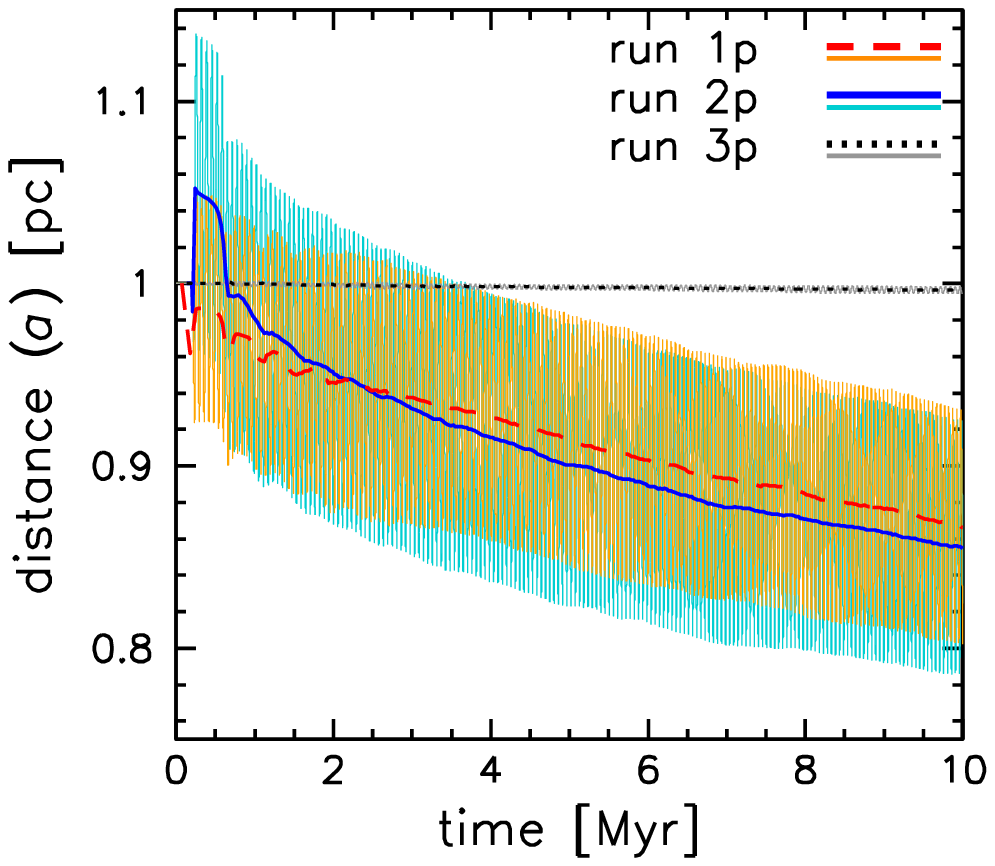} 
	\includegraphics[trim={10cm 0cm 2cm 16cm},angle=270,width=0.5\textwidth]{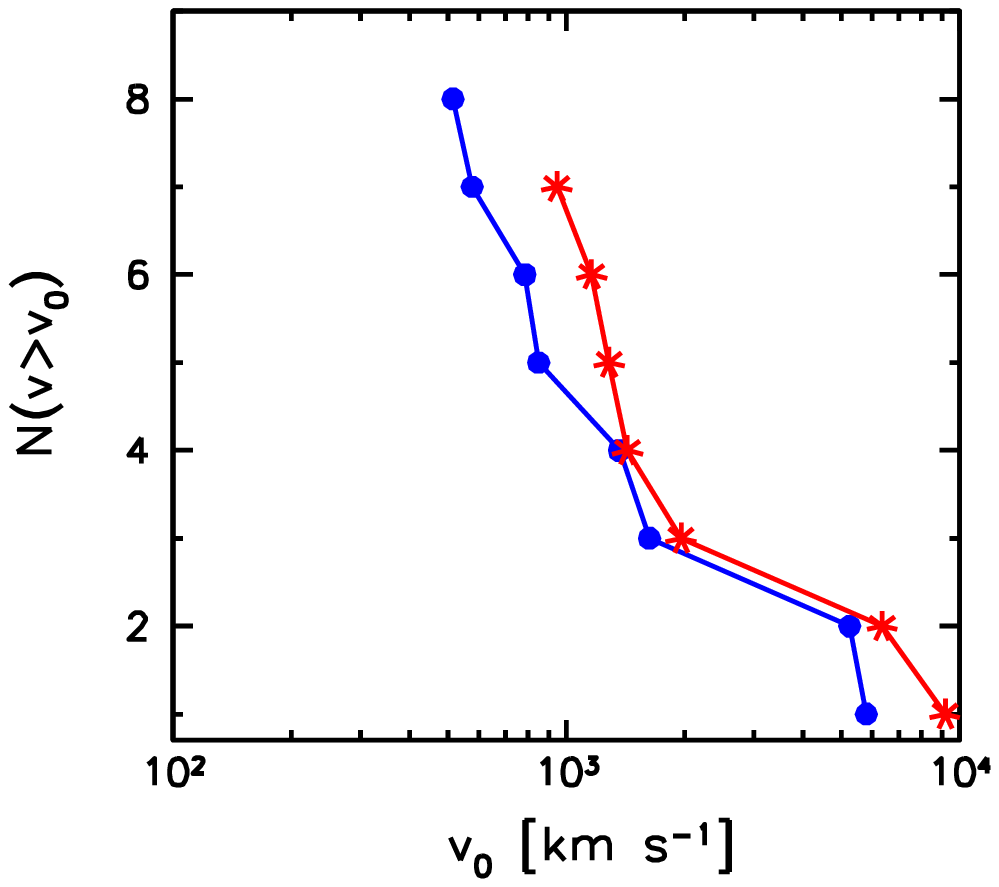} 
\caption{Left-hand panel: time evolution of the BHB separation and semi-major. Red dashed, blue solid and black dotted   thick lines show the evolution of the BHB semimajor axis  in runs 1p, 2p, 3p, respectively. The solid thin orange, light blue and grey lines show the time evolution of the BHB separation in the same runs. Right-hand panel: cumulative distribution of the number of unbound stars $N$ whose velocity $v$ is greater than a threshold velocity $v_0$, as a function of $v_0$ for run~1p (red line, asterisks) ad run~2p (blue line, circles). Note that the velocities shown here are computed from a snapshot at $t=10$ Myr. }
	\label{fig:fig2TF}
\end{figure}

If the SC is in free fall (i.e. it has zero initial angular momentum, as in runs 1p and 2p), its interaction with the BHB is rather violent: the cluster gets completely destroyed within the first $1-2$  Myr. After roughly 10 Myr, the SC stars are isotropically distributed around the BHB, and the density profile of the cluster remnant, centred on the BHB centre of mass, falls off as $\rho(r) \propto r^{-2}$ between $\approx 1 $ and 20 pc.

The time evolution of the BHB separation and semimajor axis is shown in the left-hand panel of Fig. \ref{fig:fig2TF}: after 10 Myr, the BHB has shrunk by $\approx 10$  (13) per cent in run 1p (2p)\footnote{The BHB semimajor axis, as shown  in the left-hand panel of Fig.~\ref{fig:fig2TF},  initially does not shrink monotonically: this results from the fact that the BHB is marginally soft with respect to the initial infalling SC.}; the BHB hardening is more efficient when the SC infall is coplanar, as in this configuration the relative velocity between the SC and the BHB members is lower. In the radial runs, the SC induced BHB hardening rate $s=d(a^{-1})/dt$ after 3.5 Myr is $\sim 10^{-2}$ pc$^{-1}$ Myr $^{-1}$. The BHB eccentricity $e$ is not significantly influenced by the interaction: in fact, $e<0.1$ at all times.

\subsection{SCs on non-zero angular momentum orbit}

When some angular momentum is added to the infalling SC (i.e. in run~3p, with eccentricity 0.75), the BHB response to the SC infall is significantly weaker. In particular, the BHB semimajor axis shrinks by less than 0.5 per cent in 10 Myr (Fig.~\ref{fig:fig2TF}, left-hand panel), and its hardening rate is always of the order of $s\sim 10^{-4}$ pc$^{-1}$ Myr $^{-1}$. Consistently, the BHB eccentricity stays nearly equal to zero throughout the simulation. The difference between run 3p and the other two simulations has to be ascribed to the amount of stars inside the loss cone region: in run~3p,
a negligible fraction  of stars (less than 0.3 per cent) is found to inhabit the loss cone at all times; in comparison, the fraction of stars on loss cone orbits in runs 1p and 2p is always above 25 per cent, and  higher than 50 per cent at the first SC periapsis passage.

As a result of the weak interaction, the SC stars settle on a three-lobed discy structure around the BHB (Fig.~\ref{fig:snap}, bottom right-hand panel), whose external radius is $R\lesssim 20$ pc and whose thickness is $\sim 0.1R$.

\subsection{Hyper-velocity stars}

During the interaction, most stars remain bound to the combined potential of the BHB and the galaxy throughout the 10 Myr of evolution. This happens because the BHB is still marginally soft, thus each star is expected to undergo multiple scatterings before being finally ejected from the galaxy.  However, a handful of stars manage to escape the combined potential of the SMBHs and the galaxy (7 and 8 stars, respectively) in runs~1p and 2p. The cumulative distribution of the velocity of the stellar escapers at the end of the runs is shown in Fig.~\ref{fig:fig2TF} (right-hand panel). The escape velocities at the moment of the ejection are always above 1,000 km s$^{-1}$, and can even reach 10,000 km s$^{-1}$; such escaping stars can be classified as genuine hyper-velocity stars.

\section{Discussion and conclusion}

In the present study, we studied the interaction between a $2\times 10^6$ M$_\odot$ BHB and an SC weighting roughly 1/20 of the BHB mass. We found that the interaction is rather weak if the SC initial orbit is non-radial, as no significant BHB hardening results from the SC infall. On the other hand, if the SC radially approaches the BHB, the binary semimajor axis shrinks by $\sim 10$ per cent in less than 10 Myr. 
This result suggests that the interaction of a BHB with a $\sim10$ times more massive SC (or equivalently, the radial infall of $\sim10$ SCs as massive as the ones simulated in this study) might bring the BHB close to the GW emission phase.

One may  wonder how often we expect a radial SC infall to happen. Several recent studies \cite{Tsuboi2015,Tanaka2018} show that at least part of the star formation observed in the centre of our Galaxy is consistent with being triggered by the collision between  molecular clouds. A SC formed as a result of such collision may have a very low angular momentum \cite{Mapelli2012}, thus it is expected to  infall towards the centre of its host system.

Finally, our work shows that the radial infall of an SC onto a pc scale BHB may produce a number of hyper-velocity stars, whose velocities might attain  values of the order of $10,000$ km s$^{-1}$. 

\acknowledgments{ We acknowledge financial support from the Istituto Nazionale di Astrofisica (INAF) through a Cycle 31st PhD grant, from the MERAC Foundation and from the Fondazione Ing. Aldo Gini. We also acknowledge the CINECA Award N. HP10CP8A4R and HP10C8653N for the availability of high
performance computing resources and support.}

\bibliography{biblio} 

\end{document}